\documentclass[runningheads]{llncs}
\usepackage[T1]{fontenc}
\usepackage{newtxtext}       %
\usepackage[varvw]{newtxmath}       

\usepackage{graphicx}
\usepackage{tabularx}

\begin{document}
\title{The Illusion of Relief: How GenAI Redistributes Cognitive Workload in Software Organizations}
\titlerunning{How GenAI Redistributes Cognitive Workload in Software Organizations}
%
\author{Chamari Dasanayake\inst{1}\orcidID{0009-0008-2959-8217} \and
Vladimir Mandić\inst{2}\orcidID{0000-0001-6996-2222} \and
Rahul Mohanani\inst{1}\orcidID{0000-0001-7018-8836} \and
Burak Turhan\inst{1}\orcidID{0000-0003-1511-2163}}
\authorrunning{C. Dasanayake et al.}
%
\institute{M3S Research Group, University of Oulu, Oulu, Finland \and
Faculty of Technical Sciences, University of Novi Sad, Novi Sad, Serbia
\email{chamari.dasanayake@oulu.fi}}
\maketitle              
\begin{abstract}
This study explores how Generative AI (GenAI) reshapes software practitioners' cognitive work across software development activities. We analyzed questionnaire responses from 44 practitioners across nine organizations using descriptive quantitative methods and qualitative analysis of relevant open-ended responses. Practitioners reported productivity and time savings, but also effort associated with providing context, prompting, reviewing, correcting, verifying, and integrating AI-generated outputs. Human judgment and accountability remained central, with practitioners anticipating a shift from direct code production to guiding, evaluating, validating, integrating, and coordinating AI-supported work. Thus, GenAI appears to redistribute cognitive work toward oversight rather than simply eliminate it, although whether overall cognitive workload increases or decreases remains undetermined.

\keywords{Cognitive Workload, Generative AI, Software Engineering.}
\end{abstract}
\section{Introduction}
Generative artificial intelligence (GenAI) is increasingly integrated into software development, supporting activities from requirements and design to
coding, testing, debugging, and documentation \cite{ulfsnes2024transforming,zacharias2026developers}. Although studies report productivity and efficiency benefits, particularly for routine activities \cite{ulfsnes2024transforming,brandebusemeyer2026developers}, these benefits do not necessarily imply that cognitive work disappears. GenAI-supported development can introduce effort related to prompting, evaluating, verifying, and integrating generated outputs \cite{winckler2026aiassisted}.

Prior research therefore suggests that GenAI may \emph{redistribute} cognitive work rather than simply reduce it. Effort can shift from direct task execution toward verification and supervision \cite{lee2025critical}, while software developers must review and integrate generated suggestions \cite{winckler2026aiassisted,chang2026impact}. Such changes may also reshape developer roles toward steering, reviewing, and validating AI-supported work
\cite{zacharias2026developers}. However, less is understood about how practitioners experience the movement of cognitive work across software
development activities. Examining only productivity or time savings may overlook work that emerges elsewhere in the process. The following research question guides this study---\textit{How does the use of Generative AI reshape practitioners' cognitive work across software development activities}?

We address this question through a questionnaire survey of software practitioners, combining descriptive quantitative patterns with open-ended responses.

\section{Related Work}
\label{sec:related_work}

Research on GenAI-assisted software development reports both relief and new cognitive demands. GenAI can accelerate routine work \cite{ulfsnes2024transforming}, while prompting and validating generated outputs can require additional cognitive effort \cite{winckler2026aiassisted}. In
software testing, delegating activities to GenAI can reduce effort, whereas iterative prompting and evaluation can introduce additional demands \cite{ardic2026testing}. More broadly, GenAI-supported knowledge work can shift effort from information gathering toward verification and from direct task execution toward task stewardship \cite{lee2025critical}.

Similar redistribution occurs when developers evaluate and integrate generated contributions. Effective GenAI use requires combining generated knowledge with developers' expertise and context \cite{chang2026impact}, while apparently functional generated code may still require human review and validation
\cite{glas2026insecure}. Cognitive work may therefore move from producing artifacts toward prompting, evaluating, verifying, and integrating them. The implications extend beyond individual tasks. GenAI may alter collaboration and knowledge-sharing practices \cite{ulfsnes2024transforming} and shift
developer roles toward coordination, steering, and supervision \cite{zacharias2026developers}. These findings motivate examining GenAI not only
through productivity gains, but through how cognitive work moves across the AI-supported development process.

\section{Methodology and Results}
This study employed a cross-sectional questionnaire survey to examine practitioners' use of GenAI across the software development lifecycle. The survey instrument was adapted from an earlier practitioner questionnaire developed within the same family of studies on GenAI adoption in software engineering, covering GenAI usage patterns, perceived benefits and impacts, challenges, organizational support, quality and risk management, and future expectations \cite{definex2025beyondcode}.
For the present study, the sector-specific eligibility restriction was removed, allowing software practitioners working in software development roles to participate regardless of prior GenAI use. New items were introduced on adoption hesitation, human-AI coordination, human control over AI-supported decisions, and AI-supported communication and collaboration. The questionnaire combined single-choice and multiple-selection questions, five-point agreement scales, seven-point rating scales, and open-ended questions.

The finalized questionnaire was implemented using the Webropol online survey tool. The survey was open from 28 October 2025 to 10 January 2026. Eligible respondents were current software practitioners whose professional activities involved software development. No minimum level of professional experience or prior use of GenAI was required, allowing participation from practitioners at different stages of GenAI adoption.
We distributed the survey to the participating organizations, which in turn distributed the invitation internally to employees involved in software development. Participation was voluntary and required informed consent before questionnaire access. 

Respondents could optionally provide an email address to receive results or join follow-up research; these addresses were stored separately from survey responses. Data access was restricted to the research team, and all data were anonymized before analysis. As the internal distribution process prevented determining the total number of employees who received the survey invitation, an overall response rate could not be calculated.

Responses were screened for completeness and duplication; all 44 were complete and none were excluded. Guided by the research question, the analysis examined perceived productivity and time savings, context provision and prompting, output reliability, review and validation, human responsibility, and anticipated changes in developer work. Closed-ended responses were summarized using frequencies and percentages, while relevant open-ended responses were qualitatively analyzed to contextualize these patterns. The analysis focused on shifts in the nature and location of cognitive work rather than directly measuring overall cognitive workload.

\begin{table}[t]
\caption{Indicators of cognitive work redistribution in GenAI-supported development.}
\label{tab:redistribution}

\begin{tabularx}{\textwidth}{@{}p{2.8cm}Xr@{}}
\hline
\textbf{Dimension} & \textbf{Indicator} & \textbf{Result (\%)} \\
\hline

Perceived relief
& Development process improved & 86.4 \\
& Saves at least 4 hours per week & 54.5 \\

Input and prompting
& Output quality depends strongly on input quality & 93.1 \\
& Prompting is difficult and time-consuming & 34.0 \\

Verification
& Generative AI hallucinates or fabricates facts & 63.7 \\
& Generative AI produces incorrect outputs & 54.6 \\
& Excessive iterations require more review/control & 36.4 \\

Human oversight
& Humans should remain in control of final decisions & 88.6 \\
& Developers should be accountable for AI-generated code & 72.7 \\
& Outputs are individually reviewed by experts & 52.3 \\
\hline
\end{tabularx}
\end{table}

The survey included 44 practitioners from 9 organizations, spanning diverse SE roles and averaging 14.6 years of development experience. Developers formed the largest group (19, 43.2\%), alongside practitioners in QA, architecture, management, product, DevOps, data engineering, and other roles. GenAI use was widespread---33 respondents (75\%) used it daily, 8 (18.2\%) at least weekly, and none reported never using it. Overall, 86.4\% reported improvements in their development process, while 24 (54.5\%) estimated saving at least four hours per week, indicating substantial perceived efficiency and time-saving benefits. Table~\ref{tab:redistribution} summarizes these benefits alongside additional demands related to prompting, verification, and human oversight.

\begin{figure}[t]
\includegraphics[width=\textwidth]{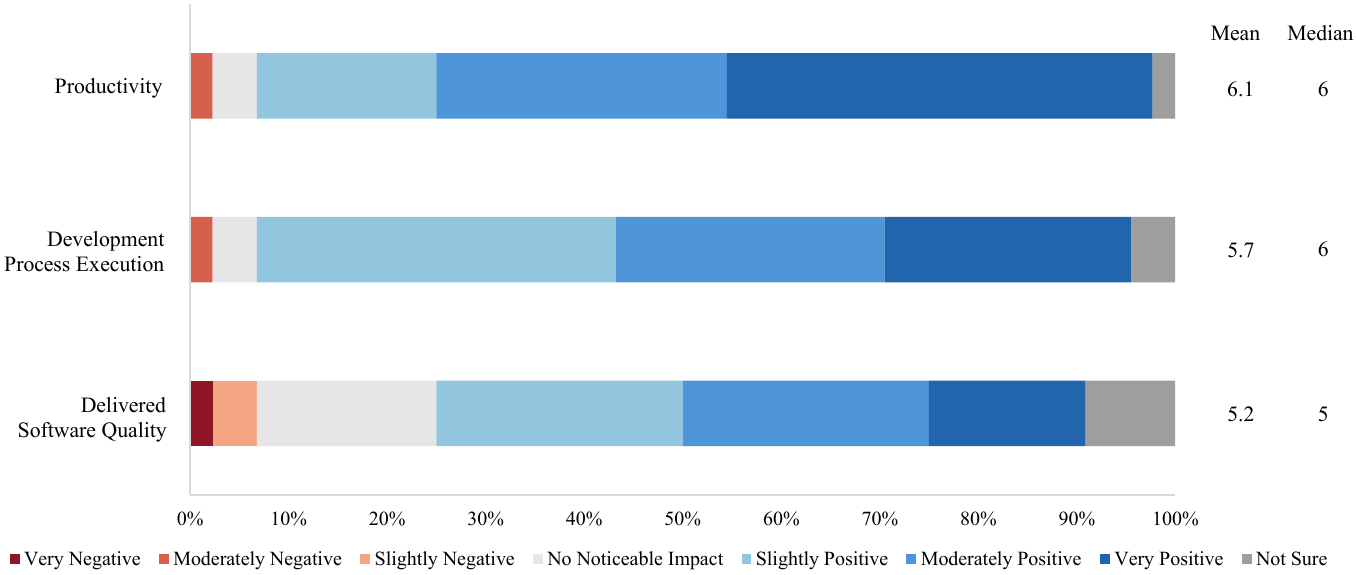}
\Description{Chart comparing practitioners' perceived impact of GenAI on productivity, software development process execution, and delivered software quality. Productivity received the most positive ratings, followed by process execution and software quality.}
\caption{Perceived impact of GenAI on productivity, software development process execution, and delivered software quality.} \label{fig:chart-impact}
\end{figure}

However, GenAI's perceived impact varied across the three outcomes
(Fig.~\ref{fig:chart-impact}). Moderately or very positive effects were reported by 72.7\% of respondents for productivity, 52.3\% for process execution, and 40.9\% for software quality, with mean ratings decreasing from 6.1 to 5.7 and 5.2, respectively. Uncertainty also increased across these outcomes, from productivity (2.3\%) to process execution (4.5\%) and software quality (9.1\%). Taken together, these results indicate that the perceived benefits were strongest for productivity and weakest for downstream software quality.

These benefits coexisted with demands elsewhere in the development process. Almost two-thirds of respondents (28, 63.7\%) identified hallucinated or fabricated facts, more than half (24, 54.6\%) reported incorrect outputs and nearly all (41, 93.1\%) indicated that output quality depends strongly on input quality. About one-third (15, 34.0\%) considered prompting difficult and time-consuming, while 16 (36.4\%) reported excessive iterations requiring additional review or control.

Open-ended responses help explain these patterns. Respondents emphasized providing sufficient system and domain context, particularly when modifying or integrating generated code. GenAI use was iterative, requiring prompting, review, correction, and evaluation before acceptance. One described a ``review-reprompt-evaluate'' workflow that can demand more effort than traditional development. Others noted that apparently functional code may contain faults or omissions, requiring developer judgment. Overall, the responses suggest that effort saved in generation may shift to context provision, reprompting, review, verification, and integration.

Human judgment remained central to GenAI-supported development: 32 respondents (72.7\%) held developers accountable for AI-generated code, 39 (88.6\%) supported human control over final decisions, and 23 (52.3\%) reported expert review of individual outputs. These findings reinforce qualitative accounts of verification and trust, showing that developers largely retain responsibility for evaluating and accepting generated outputs.

The future-role responses further illustrate where practitioners expect cognitive work to move. Respondents anticipated greater emphasis on guiding AI, reviewing and refining generated solutions, defining requirements, making architectural decisions, validating outputs, and coordinating AI agents. One respondent summarized this transition as ``more defining \& reviewing, less actual coding.'' Together, these responses indicate a perceived shift from direct production toward judgment, verification, coordination, and orchestration.

\section{Discussion}
Our findings suggest that the benefits of GenAI in software development should not be interpreted simply as a reduction in cognitive work. Respondents reported perceived improvements in productivity and time savings, consistent with prior studies showing that GenAI can accelerate routine development activities \cite{ulfsnes2024transforming} and improve perceived productivity \cite{brandebusemeyer2026developers}. However, these gains coexist with additional work required to make AI-generated contributions usable.

Respondents described effort in formulating prompts, providing context, reviewing and correcting outputs, and integrating generated artifacts. This suggests that GenAI can redistribute rather than eliminate cognitive work. Lee et al. \cite{lee2025critical} similarly identify shifts from information gathering toward verification and from direct task execution toward task stewardship. In software development, AI assistance can create a productivity-validation tension \cite{winckler2026aiassisted}, while iterative prompting and maintaining control can introduce additional cognitive demands in software testing \cite{ardic2026testing}. Our findings provide industrial evidence of this pattern: perceived relief during artifact production coexists with cognitive work related to context provision, evaluation, verification, and integration.

This redistribution also appears to change the \emph{nature} of developers' work. Respondents anticipated greater emphasis on guiding AI, reviewing and
refining generated solutions, architectural reasoning, validation, and agent coordination. This aligns with prior work suggesting a shift toward steering and supervising AI-supported development \cite{zacharias2026developers} and with evidence that effective use of GenAI requires integrating generated knowledge with developers' expertise and context \cite{chang2026impact}. GenAI may therefore reduce some production-oriented effort while increasing the importance of judgment, contextual knowledge, and evaluation.

Human responsibility remains central to this redistribution. Respondents emphasized human control over final decisions and developer accountability for
AI-generated code, while expert review was also commonly reported. This is particularly important because apparently functional AI-generated code may
contain weaknesses requiring human review and validation \cite{glas2026insecure}. Automation of artifact generation therefore does not necessarily remove human responsibility; rather, responsibility may increasingly be exercised through reviewing, validating, and deciding whether generated contributions can be accepted.

These changes also have organizational implications. Effective GenAI adoption requires more than providing access to AI tools; organizations need practices
that support review, validation, appropriate tool use, and human oversight \cite{zacharias2026developers,glas2026insecure}. As cognitive work shifts toward
evaluation and accountability, the organizational challenge may increasingly move from enabling GenAI adoption toward supporting the human judgment and assurance work that accompanies it.

\section{Conclusion}
This study examined how GenAI reshapes practitioners' cognitive work in software development. The findings suggest that GenAI provides perceived productivity and time-saving benefits that coexist with a redistribution of cognitive work from direct artifact production toward context provision, prompting, review, verification, integration, and oversight. Human judgment and accountability remain central, indicating that GenAI changes the nature and location of cognitive work rather than simply removing it.

These findings should be interpreted with the understanding that the study relies on self-reported perceptions and that the questionnaire was not designed to directly measure cognitive workload. The respondents represent nine organizations, which may limit the generalizability of the findings. The study identifies perceived shifts in cognitive work but cannot determine whether overall cognitive workload increases or decreases. Future work should examine these shifts using direct measures of cognitive workload across different development activities, roles, and organizational contexts.

\begin{credits}
\subsubsection{\ackname} This study is partially funded by the Business Finland project Managed AI-powered Software Innovations (MAISA)\footnote{https://maisaproject.eu/} (Decision number 3398/31/2024).
\end{credits}
%
%
%
\bibliographystyle{splncs04}
\bibliography{references_IS26}
\end{document}